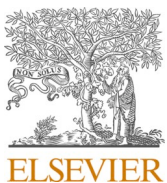

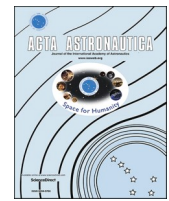

# Design and performance of a Martian autonomous navigation system based on a smallsat constellation

S. Molli [a,*], D. Durante [a], G. Boscagli [b], G. Cascioli [c,d,e], P. Racioppa [a], E.M. Alessi [f], S. Simonetti [g], L. Vigna [g], L. Iess [a]

[a] Department of Mechanical and Aerospace Engineering Sapienza University of Rome, Italy
[b] Center of Aerospace Research Sapienza, Sapienza University of Rome, Italy
[c] Center for Space Sciences and Technology, University of Maryland, Baltimore County, Baltimore, MD, USA
[d] NASA Goddard Space Flight Center, Greenbelt, MD, USA
[e] Center for Research and Exploration in Space Science and Technology (CRESST) II, NASA/GSFC, Greenbelt, MD, USA
[f] Institute for Applied Mathematics and Information Technologies "E. Magenes", National Research Council (IMATI-CNR), Italy
[g] Argotec S.p.A., Torino, Italy

## ARTICLE INFO

## ABSTRACT

Deciphering the genesis and evolution of the Martian polar caps can provide crucial understanding of Mars' climate system and will be a big step forward for comparative climatology of the terrestrial planets. The growing scientific interest for the exploration of Mars at high latitudes, together with the need of minimizing the resources onboard landers and rovers, motivates the need for an adequate navigation support from orbit. In the context of the ARES4SC study, we propose a novel concept based on a constellation that can support autonomous navigation of different kind of users devoted to scientific investigations of those regions. We study two constellations, that differ mainly for the semi-major axis and the inclination of the orbits, composed of 5 small satellites (based on the SmallSats design being developed in Argotec), offering dedicated coverage of the Mars polar regions. We focus on the architecture of the inter-satellite links (ISL), the key elements providing both ephemerides and time synchronization for the broadcasting of the navigation message. Our concept is based on suitably configured coherent links, able to suppress the adverse effects of on-board clock instabilities and to provide excellent range-rate accuracies between the constellation's nodes. The data quality allows attaining good positioning performance for both constellations with a largely autonomous system. Indeed, we show that ground support can be heavily reduced by employing an ISL communication architecture. Periodic synchronization of the clocks on-board the constellation nodes with terrestrial time (TT) is enabled through the main spacecraft (the *mother-craft*), the only element of the constellation enabling radio communication with the Earth. We report on the results of numerical simulations in different operational scenarios and show that a very high-quality orbit reconstruction can be obtained for the constellation nodes using a batch-sequential filter or a batch filter with overlapping arcs, that could be implemented on board the mother-craft, thus enabling a high level of navigation autonomy. The assessment of the achievable positioning accuracy with this concept is fundamental to evaluate the feasibility of a future positioning system providing a global coverage of the red planet.

## 1. Introduction

The growing interest in the exploration of celestial bodies of the Solar System motivates the attention towards new navigation systems, capable of effectively supporting the new scientific and commercial missions planned in the coming years.

The objective of this work is to study a navigation system consisting of a constellation of small satellites that can handle, in almost complete autonomy, the relative and absolute positioning of its nodes around planetary bodies. This system must be able to provide the necessary tools to support the navigation of other probes (e.g., objects in Entry, Descent and Landing (EDL) phase, landers, rovers, or orbiters). Using small satellites leads to low development and launch costs, short development times, and offer good flexibility in mission implementation since multiple systems can be launched simultaneously with dedicated launches

| Acronyms/abbreviations | | | |
|---|---|---|---|
| | | GPS | Global Positioning System |
| | | HPBW | Half Power Beam Width |
| ADEV | Allan Deviation | ISL | Inter-Satellite Link |
| AOCS | Attitude & Orbit Control System | LOS | Line of Sight |
| ARG | Argotec | KF | Kalman Filter |
| CDM | Code Division Multiplexing | MSC | mother-craft |
| CDMA | Code Division Multiple Access | OD | Orbit Determination |
| CCSDS | Consultative Committee for Space Data Systems | POD | Precise Orbit Determination |
| DOWR | Dual-One-Way Ranging | RTLT | Round Trip Light Time |
| DSC | daughter-craft | DSN | Deep Space Network |
| DSN | Deep Space Network | SNR | Signal to Noise Ratio |
| EDL | Entry, Descent and Landing | TT&C | Telemetry, Tracking & Command |
| EIRP | Equivalent Isotropic Radiated Power | SRP | Solar Radiation Pressure |
| GRACE | Gravity Recovery and Climate Experiment | SS | Spread Spectrum |
| GRAIL | Gravity Recovery and Interior Laboratory | USO | Ultra Stable Oscillator |
| | | WLS | Weighted Least Squares |

or as secondary payloads of larger missions.

The navigation system is based on radio observables obtained from the inter-satellite link (ISL) between the constellation satellites, heavily reducing the needs for ground support. The use of ISL technologies is widely recognized for its extraordinary results in terrestrial and planetary geodesy (GRACE and GRAIL missions) [1,2]. We demonstrate that the same concept of ISL, in the innovative configuration recently proposed [3,4] can provide observable quantities of high quality, enabling precise orbit determination, if applied to constellations with spacecraft on different orbital planes. Since the ISL provides information only on the relative motion between the spacecraft, the accurate knowledge of the Martian gravity field and rotational state acquired with previous mapping missions like MRO, MGS and Mars Odyssey [5,6], is fundamental, because it allows the conversion to absolute positioning.

We examine two different quasi-autonomous constellations of 5 SmallSats, in terms of orbital configuration and system requirements, and we compare their accuracy in terms of satellite positioning. Both constellations have a local coverage of the Mars polar regions, but this concept could be expanded to a global navigation system of the planet. The focus on the Martian polar caps origins from the need for a deep understanding of the connection between the polar deposits and the Martian climate to understand the Martian climate system [7].

The orbital period of the constellation has been selected equal to half of Mars' rotation period for the first constellation, and equal to a quarter of Mars rotation period for the second one, to have repeating ground tracks. The orbits are quasi-circular, resulting in an altitude of around 9500 km and of 4740 km, for the first and second constellation, respectively. The orbits are designed to provide navigation support during the EDL phase of a probe at high latitudes and to allow the positioning of rovers for the exploration of the Martian polar regions.

This study allows the demonstration of communication architectures potentially useful for subsequent applications, such as a future positioning system that provides global coverage of the planet.

This paper is organized as follows. In Sec. 2, we describe the design of the two constellations in terms of orbital configurations. In Sec. 3, we introduce the navigation system design based on the novel inter-satellite radio-tracking technique advised to precisely determine the orbits of the satellites, and then we focus on its implementation on the spacecraft platform. Starting from the architectural choices, in Sec. 4, we present the radio-tracking system in terms of error budgets, preparatory to the computation of the positioning performance for the two constellations. Then, in Sec. 5 we explain the orbit determination procedure, and in Sec. 6 we describe the numerical simulations in different mission scenarios for both constellations in orbit about Mars. Lastly, in Sec. 7 and 8, we show the results of the numerical simulations in autonomous and quasi-autonomous configurations to demonstrate that excellent orbital accuracies can be obtained for a constellation of SmallSat on Mars. Sec 9 gives concluding remarks.

## 2. Constellation design

This study is aimed at realizing a low-cost mission to offer a local communication and navigation system. The constellations' orbits have been designed to optimize the coverage on Mars polar regions, which are selected for scientific purposes. However, this architecture is scalable and could be expanded to realize a global navigation system or could be a part of a more complex mission concept, like MOSAIC [8].

The two constellations presented in this work are both composed of 5 spacecraft deployed on three quasi-circular, high-altitude polar orbits. The main spacecraft, the mother-craft (MSC), is placed on the central orbit, whereas the other four spacecraft, the daughter-craft (DSC), are symmetrically located on the two inclined orbits, as depicted in Fig. 1.

The orbital configuration for this navigation system arises from the initial geometric constraints:

- Orbital period as an integer submultiple of Mars' rotation period (e. g., T/2, T/4, T/8) to have repeating ground-tracks.
- Maximum Line of Sight (LOS) distance between MSC and DSCs less than 8000 km: this result from a preliminary link budget and navigation accuracy assessments. It is a value that can provide sufficient Signal to Noise (SNR) values, while offering good orbital geometry (large distances among nodes are preferred for accurate user triangulation).
- Angular separation of two couples of DSCs with respect to the MSC less than 40°: this is the assumed value for the MSC antenna's HPBW (Half Power Beam Width) for relative navigation. If the two DSCs preceding or following the MSC are always within this field of view, only two antennas are required by the MSC to perform all relative navigation links.

Consequently, the relative inclination between the MSC and DSCs orbital planes must remain in the 10–20° range, to maintain a sufficiently close distance and mitigate losses on the relative MSC-DSC links.

### 2.1. Orbital configuration

The major differences between the two configurations are the semi-major axis and the inclination of the orbits. The selected semimajor axes, 12868.6 and 8106.7 km respectively, entail an orbital period equal to half Mars' rotation period for the first constellation, hereby referred to as constellation A, and equal to a quarter of Mars rotation period for the second one, referred to as constellation B.

The orbital parameters of the spacecraft of the two constellations are reported in Tables 1 and 2, while Fig. 2 shows the spacecraft ground-tracks, which repeat every two days and one day, respectively.

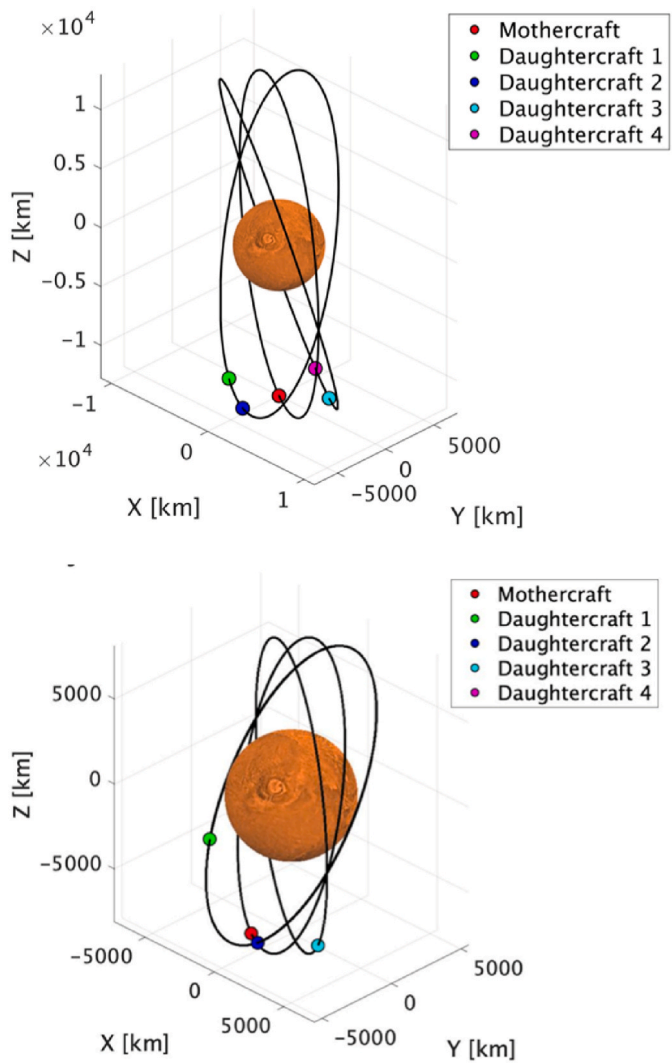

**Fig. 1.** Orbits of the constellations A (top) and B (bottom) in the IAU body-fixed frame and spacecraft relative positions.

**Table 1**
Orbital parameters of constellation A.

|                | M       | $D_1$ | $D_2$ | $D_3$ | $D_4$ |
|----------------|---------|-------|-------|-------|-------|
| **a** (km)     | 12868.6 |       |       |       |       |
| **e**          | 0.007   |       |       |       |       |
| **i** (deg)    | 90      | 75    | 75    | 105   | 105   |
| **Ω** (deg)    | −32     | −32   | −32   | −32   | −32   |
| **ω** (deg)    | 270     | 270   | 270   | 270   | 270   |
| **M₀** (deg)   | 0       | −8    | 8     | 12    | −12   |

**Table 2**
Orbital parameters of constellation B.

|                | M      | $D_1$ | $D_2$ | $D_3$ | $D_4$ |
|----------------|--------|-------|-------|-------|-------|
| **a** (km)     | 8106.7 |       |       |       |       |
| **e**          | 0.007  |       |       |       |       |
| **i** (deg)    | 75     | 60    | 60    | 90    | 90    |
| **Ω** (deg)    | −32    | −32   | −32   | −32   | −32   |
| **ω** (deg)    | 270    | 270   | 270   | 270   | 270   |
| **M₀** (deg)   | 0      | −42   | 42    | 48    | −48   |

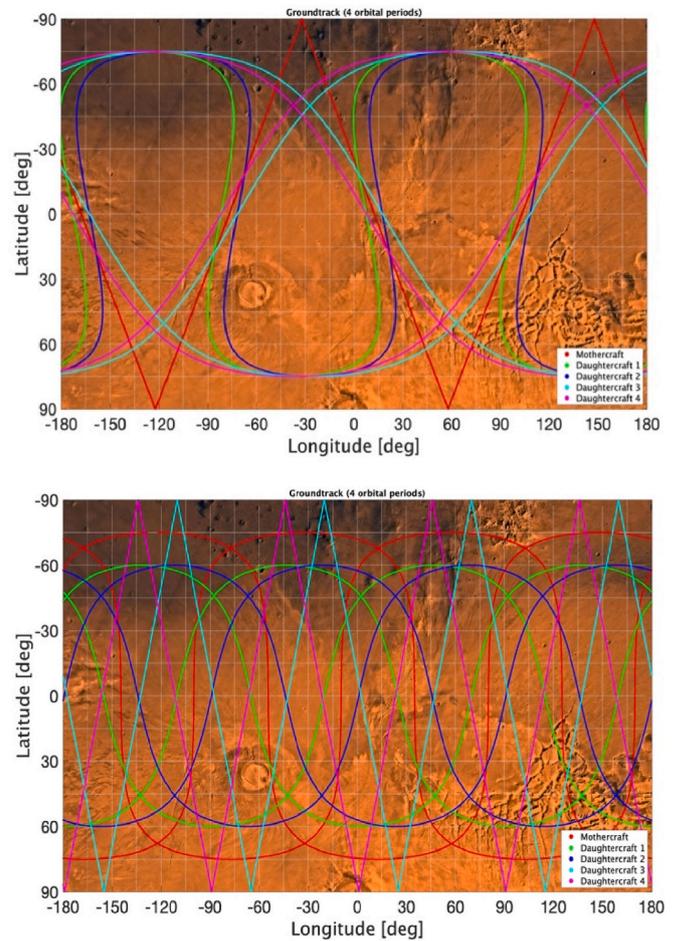

**Fig. 2.** Spacecraft ground-track constellation A (top) and B (bottom).

The maximum distance between the DSCs and the MSC is ∼4000 km for the constellation A and ∼7000 km for the constellation B.

The constellation B, differently from constellation A, is shifted in inclination by 15° with respect to the pole. This asymmetry enables a more robust position determination, leveraging the effect of the different gravity gradients.

For the same principle, the spacecraft of this constellation have a bigger baseline, given by the larger values of the mean anomaly.

To avoid collision between the satellites when their orbital planes intersect, in both constellations, the DSC are not phased equally with respect to the MSC.

## 3. Navigation system design

Deep-space ISL applications have been proposed for a variety of scenarios, spanning from the interplanetary range (e.g. Refs. [9,10]) to the atmospheric probing [11], or gravity field determination [12].

The first operative application of ISL outside of the Earth orbit has been accomplished with the GRAIL mission [2], where an inter-satellite link has been established between two spacecraft flying in precision formation around the Moon. The GRAIL concept was derived from the GRACE Earth mission and utilized a modified GRACE payload, i.e., the Lunar Gravity Ranging System (LGRS). GRAIL has obtained extraordinary results, showing that, with this architecture, the estimate of the lunar gravity field can be performed up until degree and order 1200, demonstrating robustness against both dynamic and kinematic errors [13].

These radio tracking systems were based on dual-one-way-ranging

(DOWR) measurements in Ka-band to determine the relative motion between the two orbiters [14] and required identical elements onboard the satellites. The drawbacks of this architecture lie in the large complexity and high design and development costs. Moreover, the presence of an ultra-stable oscillator (USO) on both spacecraft leads to the necessity of a precise synchronization between the two clocks, that GRACE accomplished using GPS receivers, taking advantage of being in orbit around the Earth. On the contrary, the GRAIL spacecraft required an S-band time transfer unit and a gravity recovery processor assembly (GPA) to process Ka-band and S-band radio signals that were generated through USOs. Furthermore, a daily calibration of the USO frequency was required, which was accomplished using a Radio Science Beacon (RSB) to transmit radio links at X- and S-band between the orbiters and NASA's DSN stations. The accuracy achieved in terms of range-rate was of 0.03–0.06 μm/s at 5 s integration time [15].

An alternative radio science system architecture based on range-rate measurements is presented in this work. This configuration, shown in Ref. [3], is based on a simplified scheme that leads to significant mass and power savings. In particular, in the ARES4SC study, we chose to establish cross-links only between the MSC and the DSCs, as shown in Fig. 3, realizing a "star" configuration. Recall that for *n* nodes in a constellation, the theoretical maximum number of radio links between two nodes at a time is $\frac{(n^2-n)}{2}$. Such a configuration would provide the maximum number of observables and a wider range of relative orbital geometries, which could grant higher accuracies in the determination. of the state of the constellation. However, its implementation may result in significant technical complications, and complexity in managing multiple frequencies simultaneously. For such reasons, this much simpler solution is analysed, demonstrating that it can still provide excellent positioning accuracies, while significantly containing the system complexity.

To establish the ISL between MSC and each DSC only three components are necessary (see Fig. 4):

i. An Ultra Stable Oscillator (USO) onboard the MSC, which generates the reference signal.
ii. A transceiver onboard the MSC, which transmit simultaneously the uplink signal to all the DSCs at frequency $f_1$ and then receives the downlink signal coherently from DSCs at frequency $f_2$, allowing measurement of range and range-rate.

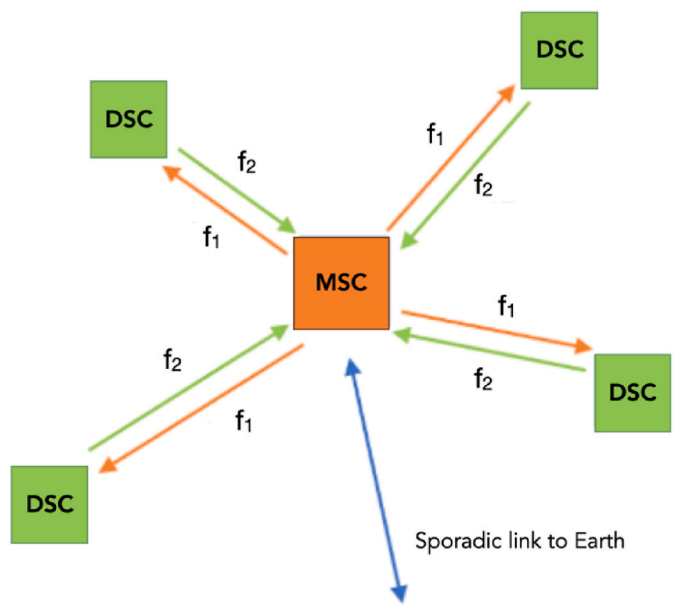

**Fig. 3.** Schematic configuration of the inter-satellite links, for a five-node constellation. Sporadic communications with ground occur through the MSC.

iii. A transponder onboard the DSCs, which receives the signal from the MSC and retransmits it back preserving phase coherency.

The selected star-like configuration, in which the observables quantities are generated onboard the MSC instead of the DSCs, minimizes the number of required USOs, allowing either the selection, for the same total cost, of a high-performance USO (in terms of Allan deviation) or the reduction of the total cost itself. To allow the use of the same frequency for all the nodes of the constellation, the use of the Code Division Multiplexing (CDM) is proposed. This choice greatly simplifies the use of the frequency spectrum, allowing an easy and simultaneous measurement of the range and range-rate, at the cost of a higher complexity of the equipment on board MSC.

The ISL is the crucial component of the autonomous navigation system, but it provides information only on the relative motion between the spacecraft. The knowledge about the absolute positioning can be possible exploiting the knowledge of the asymmetric gravitational field of the central body. If its effect is sufficiently strong, it can allow to reference the spacecraft positions with respect to the body fixed reference frame. Conversely, when the intensity of the gravitational attraction is reduced (when the orbits of the constellation are at a high altitude), an additional ground-tracking system should be provided, as it will be explained later in Section 9, allowing to reference the orbits to the Earth's surface, enabling the absolute positioning.

It must be noted that a radio link from the MSC to the Earth is anyhow required to enable Telemetry, Tracking & Command (TT&C) capabilities and to synchronize the time of the constellation with a terrestrial time scale, such as TAI or UTC. DSCs do not have such capability: the reference timescale, as well as commands from ground, are received via the MSC, which effectively acts as a relay satellite. Such a configuration simplifies the radio infrastructure on the DSCs, while retaining system complexity only on the MSC. To further simplify the overall system and reduce the complexity of the radio instrumentation, we select the X-band for ISL. This allows us to utilize the very same band for both the ISL and the link to ground, which is established in the commonly used X-band.

### 3.1. Spacecraft platform

The navigation system design of both MSC and DSC is based on the Hawk platform, which has been selected for its combination of reliability and radiation hardness that allows to deal with the harsh space environment. The Hawk platform's development, manufacturing, and qualification are carried out entirely by Argotec. Fig. 5 illustrates Argotec's HAWK-6U platform, already used for the current missions ArgoMoon and LiciaCube [16,17].

Due to the different system configurations explained later in this Section, the DSC have been designed to be smaller platforms (6U CubeSat) with respect to MSC (12U CubeSat). Only the 12U CubeSat can take the MSC role in the selected constellation system, leading to a simpler radio link configuration. The constellation can be designed to add redundancy to the whole system considering that some (or all) DSC could be replaced by 12U CubeSat with the possibility that any of these satellites can take the MSC role.

The MSC is equipped with a navigation payload that is responsible for the provision of the autonomous navigation service. It is composed by:

• **Ultra-Stable Oscillator (USO):** the Ultra-miniature OCXO 9700 from Microsemi [18] has been preliminarily selected. It has stability in terms of Allan Deviation (ADEV) of $5 \bullet 10^{-12}$ approximately flat from 1 s to 1000 s, an acceptable value to obtain good accuracies in the orbit determination and time synchronization process, (approximately, two order of magnitude better at 100 s w.r.t. the typical ADEV of a X-band transponder with its internal oscillator (OCXO) in non-coherent mode), and low mass and volume requirements.

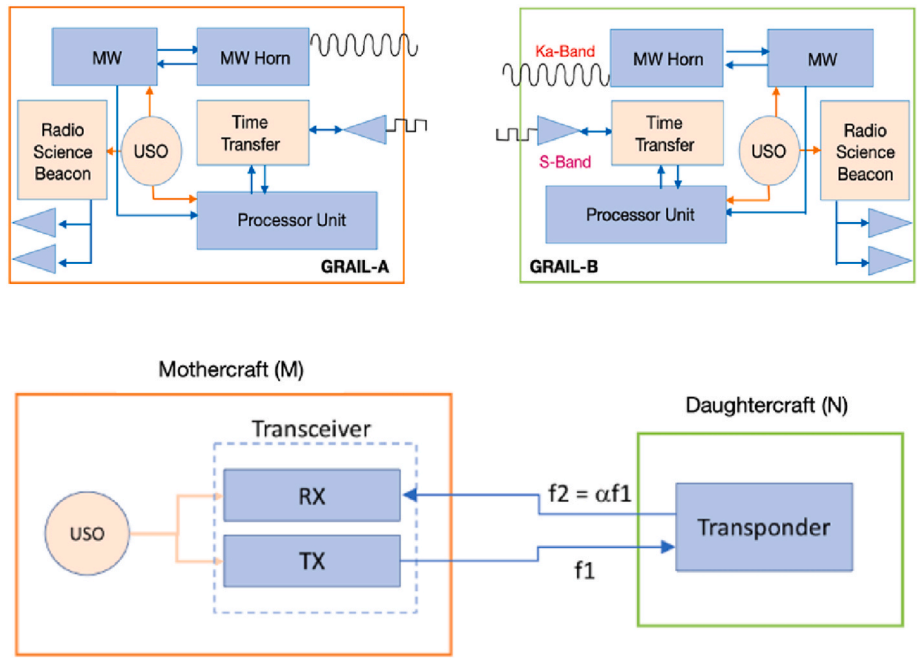

**Fig. 4.** Block diagram of GRAIL (top) [2] and the novel (bottom) radio system for intersatellite tracking [3].

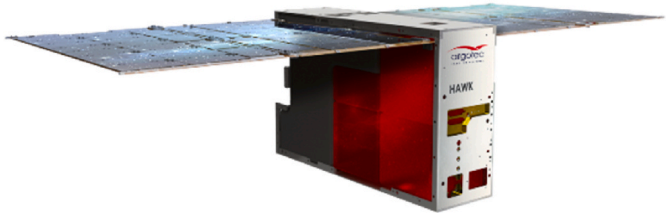

**Fig. 5.** Argotec's HAWK-6U platform.

- **Two Tx/Rx Antennas:** given the orbital configuration, two antennas, facing opposite sides, are needed to allow simultaneous communication with all the DSC. An ISL budget and coverage analyses have been performed to select the antennas. A couple of TX/RX patch antennas have been identified with a gain at boresight >11.5 dBi between 8 GHz and 8.4 GHz and an HPBW of 40 deg (which is in line with the maximum angular separation of two DSC).

- **X-Band Transceiver:** The transceiver shall support, for the simultaneous communication with all the four satellites of the constellation, a spread spectrum (SS) multichannel architecture based in transmission on a CDM-M scheme and in reception on a classical parallel SS receiver (CDMA solution) [19]. The transceiver shall be compliant with CCSDS Standard and SFCG Recommendation for communication in the Mars Region [21]. The minimum RF power (4 W) is derived from a preliminary link budget analysis, and it is limited by the platform power availability. Moreover, to transmit the same signal to all the DSCs nodes, a power splitter is needed to redirect the signal to both Tx antennas. A compatibility analysis and a trade-off have been performed to select the transceiver. A lack of availability of commercial telecommunication payload, for CubeSat, supporting CDMA scheme has been identified. A Software Defined Radio has been preliminarily considered; it is partially compliant with the requirements, and it is well suited for a future reconfiguration to support CDMA communications and generation of navigation signal.

The DSC is equipped with a navigation payload composed by:

- **Tx/Rx Antenna:** it provides the RF signal reception and transmission from/to the MSC via one couple of patch antennas onboard the DSCs (which are the same as the MSC) pointed toward the mother node.

- **X-Band Transponder:** it provides the coherent retransmission of the ranging signal sent by the MSC. To be fully compliant with the CCSDS standard, the signal shall be regenerated meaning that the transponder onboard the DSC shall support CDMA technique too. For this reason, in this preliminary selection of the transponder, a high priority is given to mass and volume requirements and reconfigurability. The Iris V2 [22], a miniaturized transponder of limited mass, has been successfully used on board the MarCO SmallSats at Mars [23], and it's used for the current missions ArgoMoon [16] and LICIACube [17]. Future reconfiguration to support CDMA communications must be considered also in this case.

With the chosen antennas, the ISL budget, at the maximum distance of 7000 km shows, for the forward link, a SNR of 30 dB-Hz, with a command data rate in the range of 200 bps; for the return link a SNR of 35 dB-Hz, with telemetry data rate in the range of 600 bps; for the ranging measurement a rms value of 1.25 m at integration time of 1 s.

To communicate with ground (downlink), high-gain antennas onboard the satellite are needed. Assuming an X-Band communication with 35 m antennas on the ground, it results that to ensure a downlink data rate ranging from 250 bps (2.65 AU) to 8 kbps (0.5 AU) in a nominal scenario, an Effective Isotropic Radiated Power (EIRP) greater than 35 dBW is needed. For this reason, the platform configuration foresees a reflect array antenna with an aperture size of 0.60 × 0.33 m, a gain of 29.2 dBi and an HPBW of 4.9 deg [24,25]; a transmission RF power of 6 dBW is needed to satisfy the requirement.

The allocation of the reflect array on the MSC has been made possible by exploiting an integrated reflect array and solar array panels which is a technology demonstrated in both ISARA [26] and OMERA [25] missions.

The link from ground to the MSC (uplink) is possible with a medium-gain antenna (2 × 2 patch array) having a gain of 12 dBi and an HPBW of 40 deg.

For constellation redundancy purposes, also the DSCs can communicate with Earth at a low data rate only in case the MSC is lost. The same

DSC transponder, used for navigation, can be used for navigation with Earth. Two antennas have been mounted on one side of the spacecraft for TT&C. The Rx antenna is an X-Band 2 × 2 patch array antenna, instead, the Tx antenna is an X-Band 4 × 8 patch array antenna with a gain of 22 dBi.

In addition to the Navigation payload and TT&C, both MSC and DSC are composed of the following subsystems: Structure, Thermal Control, Electrical Power (i.e. battery, Solar Panel Array and Power Conversion and Distribution Unit), Attitude Determination and Control (to determine and control where the satellite should be pointed and orient it to establish Inter-Satellite Link, link with Earth, to perform sun-pointing for battery charging, and reaction wheels desaturation), On-Board Computer and Data Handling (the selected one is the OBC&DH Mk I produced by Argotec with heritage from both the ArgoMoon and the LICIACube deep-space CubeSats), On-Board Software, Propulsion and Harness.

Fig. 6 and Fig. 7 show the layout of the MSC and DSC platforms, respectively, including the main external equipment location, namely the navigation antennas, the antennas for the link with Earth for TT&C, and the Solar Panel Array.

## 4. Radio-tracking system

This section describes the error budgets for both Earth link and ISL Doppler observables.

### 4.1. Space-to-ground

The key parameter determining the achievable accuracy of the Doppler tracking system is the frequency stability of the GS–SC–GS loop. Typical performance of deep space Doppler tracking systems employing Doppler measurements (here reported at 10 s integration time) is of the order of 0.2 mm/s for X-band radio links [27] and 0.05 mm/s at Ka-band [28], reaching levels up to 0.015 mm/s for state-of-the-art systems such as the one of BepiColombo [29].

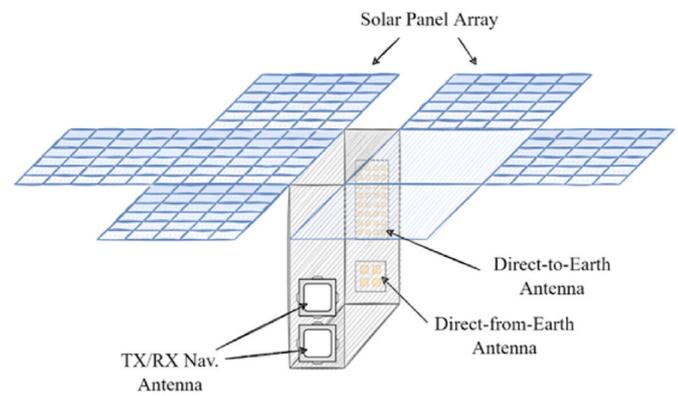

**Fig. 7.** DSC – 6U CubeSat layout showing the main external equipment.

Non-signal disturbances in a Doppler link are due to instrumental noises (random errors introduced by ground or spacecraft themselves), propagation noises (random frequency/phase fluctuations caused by refractive index fluctuations along the line of sight), or systematic errors. However, analyses and experience have shown that at timescales ~1–1000 s systematic contributions are negligible.

The radio tracking data at X-band are mainly perturbed by interplanetary plasma. A plasma model proposed by Iess et al. [30] provides the Allan deviation as a function of SEP angle, which is strictly valid for missions in the outer solar system, showing that the level of plasma noise increases approaching superior solar conjunctions. The noise caused by fluctuations of the tropospheric refractive index is another major contributor. At microwave frequencies, tropospheric refractive index fluctuations are nondispersive and dominated by water vapor fluctuations [31]. Up to ~90% of the tropospheric path delay is due to the dry, or homogenous, part of the troposphere. The induced Doppler signal in ~6 h of deep space tracking amounts to ~1 mm/s, but can be efficiently resolved with suitable elevation models and ground readings of meteorological data. On the other hand, the wet path delay amounts to only a few centimeters, but its fluctuations are hardly predictable because of the non-homogenous distribution of the water vapor. Finite SNR of electronic devices on the radio links, on both the ground and spacecraft equipment, induce phase instabilities on the signal carrier. Noise contributions due to the frequency standard and electronics of the ground station are usually negligible. The main contribution from electronic systems comes generally from the onboard deep space transponders (DSTs), unless special design is adopted (as for Juno and BepiColombo Ka/Ka radio science transponder). These devices usually exhibit a white phase noise ($S(y) \propto f^2$) at frequencies larger than $10^{-4} \div 10^{-3}$ Hz. Transponders based on digital architectures generally have poorer stability, like Rosetta's DST, with Allan deviations $\sigma_y \approx 4.7 \cdot 10^{-14}$, or 0.015 mm/s range rate accuracy, at 60 s integration time. For comparison, Cassini's analog DST performs better ($\sigma_y \approx 1.8 \cdot 10^{-14}$ or 0.006 mm/s at the same integration time) [30]. Thus, noise contributions due to the last generation of digital transponders are negligible too. A significant contribution between instrumental noises to the overall Doppler error budget comes from thermal and mechanical deformations of the ground antennas. These deformations are caused by wind and gravitational loading, and time-varying thermal gradients across the antenna structure. An analysis performed by ESA showed that antenna mechanical noise at ESTRACK antennas is at level of $\sigma_y \sim 1.6 \cdot 10^{-14}$ (0.05 mm/s) at 60 s integration time [32]. Ground antennas' thermo-mechanical stability is well below other noise sources, but not negligible.

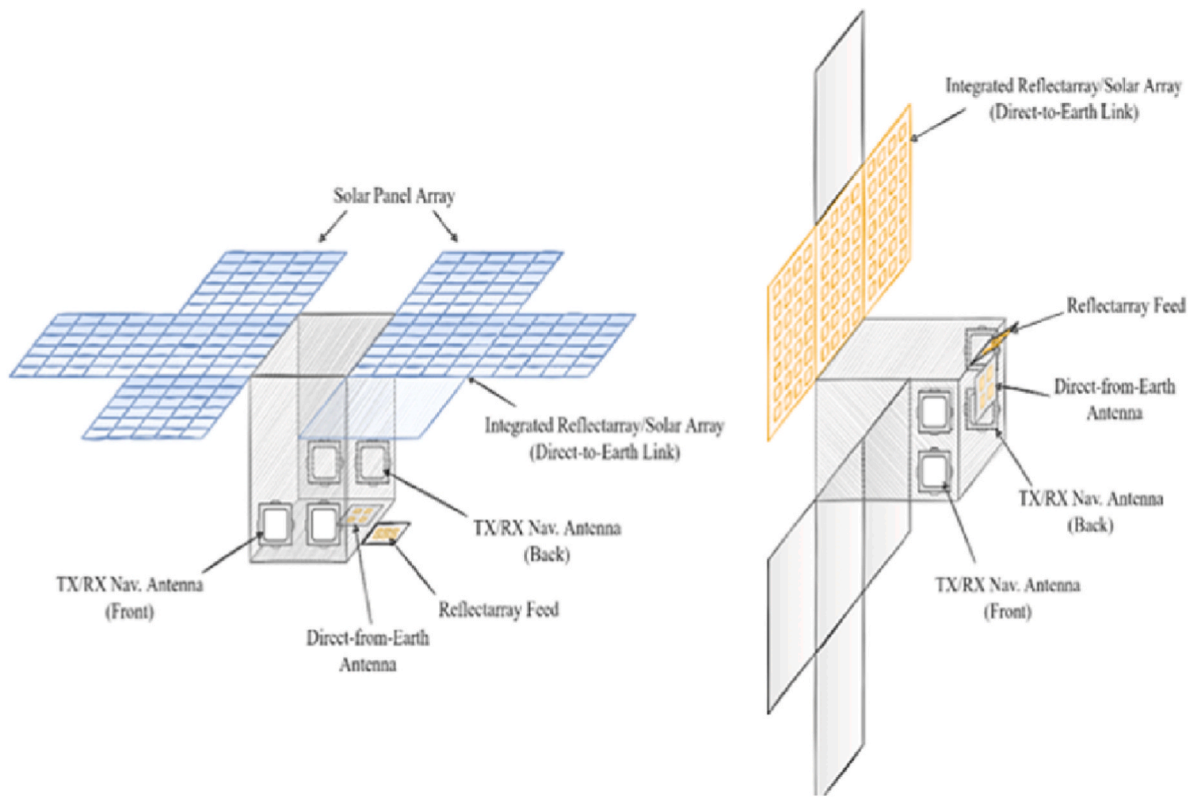

**Fig. 6.** MSC – 12U CubeSat layout showing the main external equipment.

Consequently, the error budget for space-to-ground Doppler measurements indicates an accuracy in range rate measurements of 0.04 mm/s @ 60 s in X-band.

### 4.2. Inter-satellite link

The main factors that determine the Doppler accuracy are:

- the error caused by thermal noise of the receiver;
- the stability of the transponder on board the nodes;
- the stability in frequency (ADEV) of USO.

As stated in Section 2, the maximum distance between the nodes was selected as to have adequate power levels, limiting the contribution of the thermal noise, that could be expressed in terms of ADEV as follows [33]:

$$\sigma_y(\tau) = \frac{1}{2\pi f \tau} \sqrt{\frac{3B_l}{SNR}}$$

Assuming the noise bandwidth of the receiver ($B_l$) equal to 0.1 Hz, considering the worst-case scenario with an SNR equal to 30 dB-Hz, the thermal noise contribution at X-band ($f = 8.4$ GHz) is $3.3 \times 10^{-16}$ at $\tau = 1000$ s and scales with $\tau^{-1}$.

The latest generation transponders for deep space applications showed ADEV of the order of $3–4 \times 10^{-16}$ at 1000 s integration time and about $1–2 \times 10^{-15}$ at 60 s. In this study we will assume ADEV equal to $1 \times 10^{-14}$, comparable with that of Iris V2. In terms of *range-rate error (2-way)* this frequency stability corresponds to 3 μm/s at 60 s of integration time [22].

The frequency stability of the USO is the fundamental element to ensure the accuracy of range rate measurements. As demonstrated in Ref. [3], in a two-way Doppler link, the limited spatial separation, therefore the short round-trip light time (RTLT), of the order of 10–20 ms at most, implies a strong suppression of the noise of the USO. In fact, since the same reference frequency is used in both uplink and downlink, any instability that appears at time scales larger than the RTLT is strongly reduced. This cancellation effect can be appreciated considering that the clock noise transfer function, in a 2-way link, consists of two anticorrelated Dirac delta ($\delta$ (t)) functions separated by a RTLT. Let us call $y_c$ the random noise process associated with the relative frequency fluctuations introduced by the local oscillator used to generate the reference signal. The instability in the relative frequency shift $y(t) = \frac{\delta f}{f}$ between the received signal and the reference signal can be written as [3]:

$$y(t) = y_c \delta\left(t - \frac{2L}{c}\right) - y_c \delta(t) =$$

$$yc \, \delta(t - RTLT) - y_c \delta(t)$$

where L is the spatial separation between the two nodes. The anti-correlation at one RTLT leads to strong noise suppression when the integration time T is much greater than the RTLT. In fact, for low frequencies, the corresponding noise power spectrum $S_c(f)$ of the USO becomes:

$$S_y(f) = 4S_c(f)sin^2\left(\frac{\pi fL}{c}\right) \approx 4\pi^2 S_c(f)\left(\frac{fL}{c}\right)^2, \frac{\pi fL}{c} \ll 1$$

By integrating over a time T (typically 60 s), the spectral power at frequencies higher than $\frac{1}{T}$ is attenuated by a factor proportional to:

$$4\left(\frac{fL}{c}\right)^2 \le \left(\frac{2L}{Tc}\right)^2 = \left(\frac{RTLT}{T}\right)^2$$

For the ADEV, the suppression factor is equal to its square root:

$$\sigma_y(T) \propto \sigma_{yc}(T)\frac{RTLT}{T}, \quad \frac{\pi \, RTLT}{T} \ll 1$$

Where $\sigma_y$ is the ADEV of the link measured by the transceiver if only the noise of USO were present and $\sigma_{yc}$ is the ADEV of the USO.

Considering the worst case with 7000 km of inter-satellite distance and T = 60 s, the noise of USO is therefore attenuated by a factor of approximately 800. With the chosen USO with ADEV equal to $5 \times 10^{-12}$, the residual clock noise in the Doppler measurement would be equal to $6 \times 10^{-15}$, corresponding to an error in range rate equal to 2 μm/s. This configuration, therefore, allows to obtain excellent range-rate measurements with a very cheap and light USO. The error budget indicates a worst accuracy in range rate measurements of 3.9 and 3.6 μm/s at 60 s integration time for constellation B and A, respectively, as shown in Fig. 8.

As a comparison, Ka-band range-rate (KBRR) measurements from GRAIL mission [2], directly derived from the KBR data, have an accuracy of 0.03–0.06 μm · s⁻¹ at 5 s integration time but with a very sophisticated system, highly demanding in terms of mass, power, and volume requirements.

## 5. Orbit determination procedure

To investigate the performance of the mission concept ARES4SC, the first step of the orbit determination (OD) procedure is the generation of realistic data from the s/c, called *synthetic observables*. These are collected using a *simulator* that obtains a reference trajectory propagating the orbits starting from the chosen initial conditions, using a provided dynamical model. Then, an observational model and a realistic noise process are implemented to simulate the observables. The key part of the OD procedure is the *estimation*, in which the differential corrections procedure is applied from the available observation to solve for the parameters we are interested in, called *state vector* [34]. This can be solved with two class of estimators: *batch* and *sequential*. The key difference is that batch estimators wait for all observables to be collected and use all of them in a single inversion to estimate the state vector at the desired reference epoch, while the sequential algorithm process one observation (or a few) at a time. One of the most used batch estimators is based on a linear *least squares* formulation. For this work, we adopt a filter based on optimally weighted least square (WLS hereafter, see 5.1 for a discussion on the weighting procedure) with a-priori information. Due to the high non-linearity of the OD problem, the inversion process shall be iterated a few times to reach convergence. Nevertheless, the solution obtained with this method could not be sufficient when the dynamics of the spacecraft is quite complicated, because the mathematical model and/or the numerical integrator may not be accurate enough for very long data arc, leading to an integrated trajectory that is not consistent with the observations. This is mainly related with difficulty in modelling some of the forces, as in the case of Solar Radiation Pressure (SRP) and thruster firings for trajectory and attitude controls. Furthermore, in several space geodesy experiments the number of *solving-for* parameters is quite large. This problem is in many cases severe, to the point that a single batch estimation becomes impractical. In such case, the trajectory could be subdivided into different non-overlapping *arcs* (batches). At this point, different batches can be combined to obtain a unique, and better, estimate of a common subset of parameters, or consecutive batches can be combined sequentially to update the state estimate for the same state vector. In the former approach, called "*multiarc*" [35], the common "*global*" parameters (e.g., gravity field coefficients of a body) are uniquely solved, while the non-common "*local*" parameters (usually the initial position and velocity components) are replicated independently for each arc. The second approach allows performing a near-real-time OD of the constellation with the more robust and accurate batch method. In this case, batch filtering (with few iterations) of a generalized state vector for the whole constellation is solved in each arc. The different estimates may be used in combination with different approaches depending on the problem. Batch methods are, in general, computationally more demanding (both for the iterations and the size of involved vectors and matrices) than sequential methods and are not well suited for real-time applications. The latter is the typical area of applicability for sequential estimation methods, providing a state estimate at the observation epoch that is continuously updated after each (or a few) new observation is collected. Kalman Filter (KF) is an example of Sequential filter, linear as well, computationally light (mainly because of the lower memory requirements), but it is not stable enough for OD applications. Extended Kalman Filter (EKF), or Unscented Kalman Filter (UKF) are better options (although computationally more demanding), for real-time OD. In this work, we investigate the first segment of the ARES4SC mission concept: the numerical warm-up period necessary for the constellation to assess its ephemeris, analysing the obtainable accuracies to understand if such a concept is feasible for a future navigation system. Therefore, real-time application is not required for this purpose, leading us to the choice of batch filters, mainly for its robustness. We propose two different procedures for this approach: a first one, called *batch sequential filter*, in which the components of the spacecraft position and velocity vector, called *propagated parameters,* and their covariance, estimated at the beginning of the arc, are propagated until the beginning of the next arc, and used as a priori information, optionally using a conservative scaling factor. It's also possible to set constant parameters, such as the gravity field coefficients or the central body's state vector at a given reference epoch, as *local parameters*. In the second procedure, called *batch with overlap filter*, the covariance of the propagated parameters is not mapped, since a priori values for each parameter are fixed at the beginning of the process for all the arcs to non-constraining values. Furthermore, to have a smoother solution, the data batches are overlapped. A simplified scheme of the implementation of both filters is provided in Fig. 9. In both approaches, each data arc is solved with WLS filter, which is the building block of the POD algorithms. Future works will assess if the same methods could be used also during the user navigation phase of this mission or if a sequential filter will be needed. This consideration will depend mainly on the quality of the *ephemeris ageing* along the length of the arc.

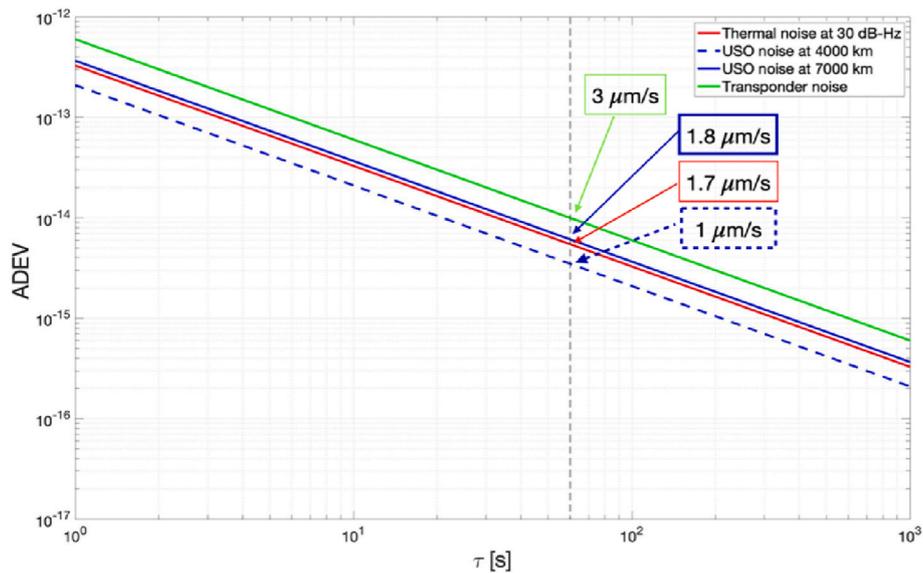

**Fig. 8.** Error budget for the ISL. The USO noise is represented at 7000 km and 4000 km (maximum distances for constellation B and A, respectively), and thermal noise at SNR = 30 dB-Hz (reference value from link budget).

### 5.1. Weighted least squares with a priori information

The OD problem is heavily non-linear as both the state vector $X$ and the observation vector $Y$ are described respectively by a set of $n$, the number of parameters to estimate, first order non-linear equations $F(x, t)$, and $m$, the number of observables in the batch, non-linear equations $G(x,t)$.

In order to apply the least-square principle, that is a linear estimator, the problem shall be linearized expanding around a reference trajectory $x_{ref}$, and truncating at the first order. The problem is then reformulated in terms of the state deviation vector $x$ relative to the reference trajectory, and the observation deviation vector $y$ describing the difference between the observed observables and the computed observables on the reference trajectory.

This produce the differential corrections estimation method where the governing equation are given by:

$$\dot{x}(t) = A(x,t)x(t) \quad y(t) = \widetilde{H}(x,t)x(t) + \varepsilon(t)$$

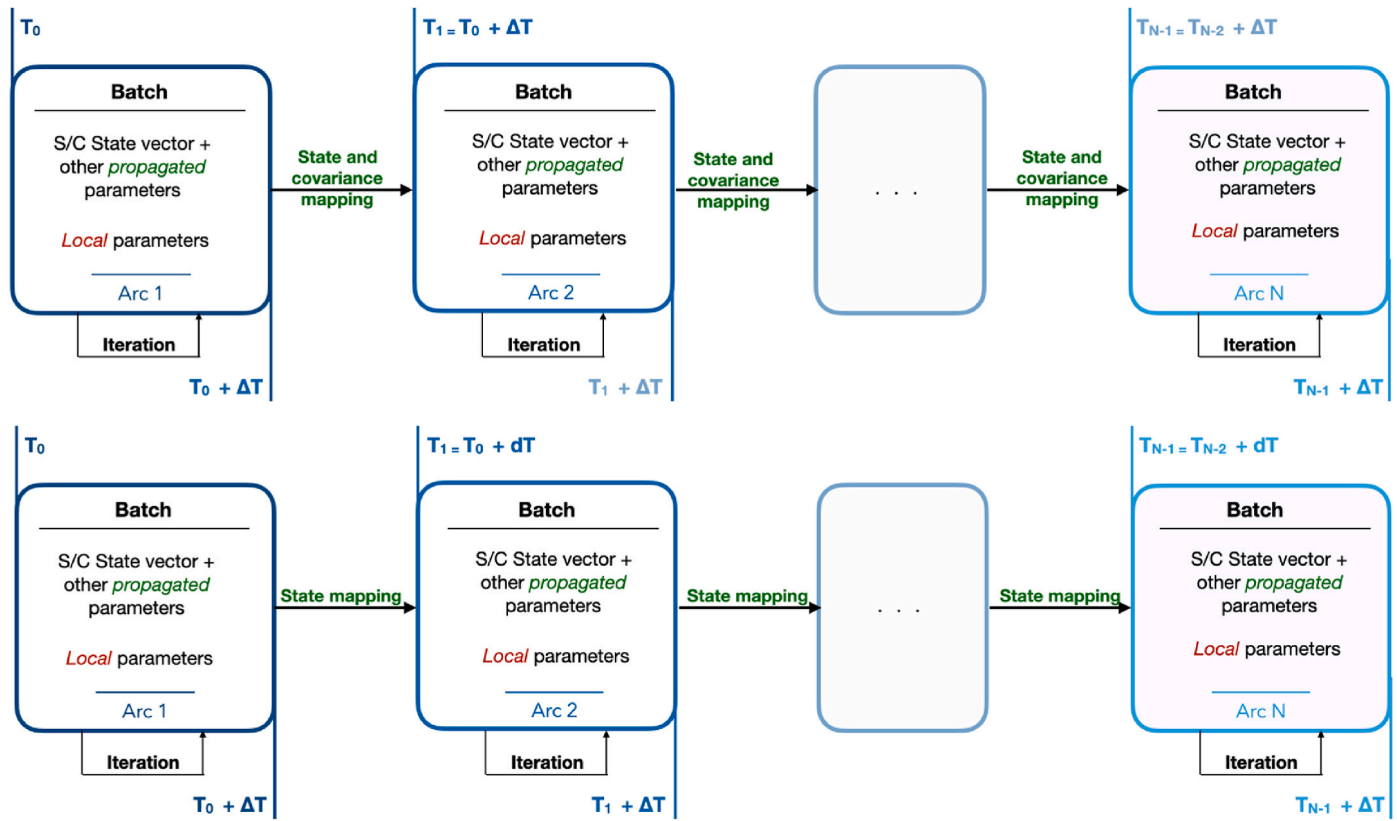

**Fig. 9.** Batch-sequential scheme (top) and batch filter with overlap (bottom).

Where:

$$A(x,t) = \frac{\partial F(x,t)}{\partial x(t)}\Big|_{x=x_{ref}} \quad \widetilde{H}(x,t) = \frac{\partial G(x,t)}{\partial x(t)}\Big|_{x=x_{ref}} \quad \varepsilon(t) \sim N(0,R)$$

Here $x(t)$ is the unknown state deviation at the observation time $t$. There is a different state deviation for each set of observables at time $t$. Instead, $\varepsilon(t)$ is a random vector of size $m$ accounting for the observation errors due to measurement noise, assumed to be gaussian with zero mean and covariance matrix R. The matrices $A$ and $\widetilde{H}$ contains partial derivatives computed along the reference trajectory for the dynamical and the observations model respectively.

Note that $\widetilde{H}(x,t)x(t)$ is the correction term accounting for all the modelling and deterministic observation errors. Indeed, we are explicitly assuming the noise is zero mean, so that the estimate is unbiased. This means that any bias in the collected observations shall be calibrated a priori or estimated within the state vector.

The key step in a batch filter is to relate all the $y(t)$ to the same state deviation $x_0$ at time $t_0$. For this, the state transition matrix $\Phi(t,t_0)$ is used to multiply to produce each row of the design matrix $H$ so that a single observation deviation equation can be written for the full batch of observables (the vector $y$ and $\varepsilon$ are now not referred to a specific time):

$$y = Hx_0 + \varepsilon$$

Where:

$$x(t) = \Phi(t,t_0)x(t_0) \quad H = [\widetilde{H}(x,t_1)\Phi(t_1,t_0) \vdots \widetilde{H}(x,t_m)\Phi(t_m,t_0)]$$

The state transition matrix is obtained by solving the linear differential equation $\dot{\Phi}(t,t_0) = A(x,t)\Phi(t,t_0)$ with the initial condition $\Phi(t_0,t_0) = I$. The obtained equation is also known as the batch data equation (a linear vector equation of size $m$). We look for $x_0$, the differential correction to the state vector relative to the reference trajectory of the current iteration, by minimizing the sum of squares of the residuals $\varepsilon(t)$, so that we can correct for the wrong values in the reference state vector at time $t_0$ and obtain post-fit residual containing only the unavoidable observation noise.

The a priori information are included considering an equivalent data equation that relates the a priori knowledge on the state vector deviation relative to the same reference trajectory as above. That is a data equation were $H = I$ and the associated noise covariance is the a-priori state covariance matrix $\underline{P}$:

$$\underline{x} = x_0 + \varepsilon(t)$$

Where:

$$\nu(t) \sim N(0,\underline{P})$$

Here $\underline{x}$ is the a priori state deviation relative to the reference trajectory at the estimation epoch $t_0$. It is usually zero for the first iteration, when the same reference trajectory and model used to linearize the problem is also that providing the a priori information. On subsequent iterations instead, this it will be different from zero as we will use the updated trajectory as new reference point, but a priori information does not change. If we call $\hat{x}_k$ the estimated state deviation at iteration $k$, the next a priori state deviation will be:

$$\underline{x}_{k+1} = -\sum_{i=1}^{k} \hat{x}_i$$

Introducing the $m$-by-$m$ weighting matrix $W$, the least squares solution is obtained by minimizing the scalar cost function C, given by the sum of squares of the observed and the a priori residuals:

$$C = (y - Hx)^T W(y - Hx) + (\underline{x} - x)^T \underline{P}^{-1}(\underline{x} - x)$$

The solution for the estimated state deviation $\hat{x}$ and estimated state covariance $\hat{P}$ is:

$$\widehat{x} = \widehat{P}\left(H^T W y + \underline{P}^{-1}\underline{x}\right) \quad \widehat{P} = \left(H^T W H + \underline{P}^{-1}\right)^{-1}$$

## 6. Numerical simulations

In our numerical simulation, we have created synthetic observables by adding a realistic measurement noise for ISL and ground-tracking based on the error budgets shown in Section 4. As shown, the ISL radio link accuracy depends on the inter-satellite distance, while the Earth-link accuracy have been set to 0.04 mm/s @ 60s, all in X-band.

The error budget of the pseudolites links has not been investigated in this study, where we conservatively assumed the same accuracy of the Earth link.

To estimate the positioning performances of the constellation nodes, we exploited POD algorithms based on the two filters described in Section 5.

The dynamic equations of the simulated trajectories account for the Mars gravity field expressed in spherical harmonics to degree and order 120 (MRO120D [5]), the gravitational attraction from Solar system planets and the Martian satellites in a relativistic context, and the non-gravitational acceleration caused by Mars albedo, its infrared radiation, and SRP on the satellites.

The spacecraft shape and thermo-optical properties are based on the SmallSat design being developed in Argotec and described in Section 3.

Different mission scenarios were investigated to compare orbital solutions obtained through the following analysis:

1. Intersatellite radio only, simulated 24/24 h;
2. Alternating combination of ISL and deep-space radio link;
3. Alternating combination of ISL and pseudolites link.

Numerical simulations of these nominal scenarios were carried out with ESA Godot software to investigate the expected positioning accuracies.

We performed a perturbation analysis for both constellations, comparing the positioning performance achievable with the two estimation filters we selected. The choice of the batch duration (arc length) is particularly important for a fast convergence and a good accuracy of the final orbital solutions. Its length is a trade-off between the complexity of the estimation procedure (amount of data to be processed in a batch) and the estimation accuracy (the more data processed, the better accuracies). The analysis was carried out for a mission duration of 24 days, with an arc length of 4 days for the batch-sequential filter, while for the batch filter we considered batches of 7 days with 4 days of overlap.

We perturbed the dynamical model of the estimation process by limiting the spherical harmonics expansion of Mars gravity field up until degree and order 20 (to limit the computational cost on-board the satellite) and modified the thermo-optical coefficients of the spacecraft shape elements as to introduce a ~10% mismodeling error on both the SRP and the Mars albedo accelerations. Hence, we added a set of white-noise stochastic accelerations along the Radial, Transverse and Normal (RTN) directions as estimated parameters (*local parameters*) to absorb the errors. We selected a time update of 3 h after an accurate analysis of the positioning performance obtained considering a wide spectrum of values.

## 7. Autonomous system results

We report in Figs. 10 and 11, the estimation error, i.e., the difference between the estimated and simulated trajectories, the 1-σ and 3-σ uncertainties for the case with ISL alone, of the R, T and N components for the two constellations using both filters.

This analysis shows, in the case of ISL-only and constellation A, that the trajectories cannot be recovered with confidence (the error in the normal component is larger than formal uncertainty). As shown in

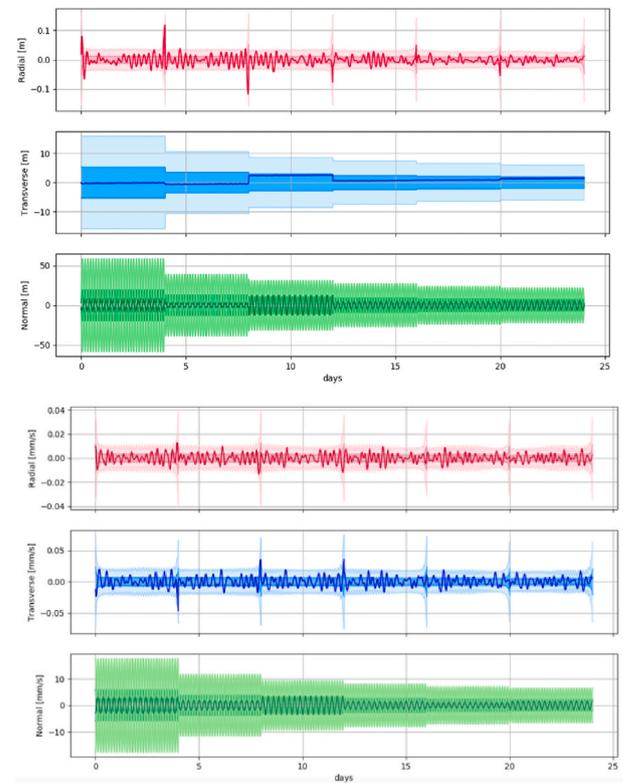

*a) Batch-sequential filter*

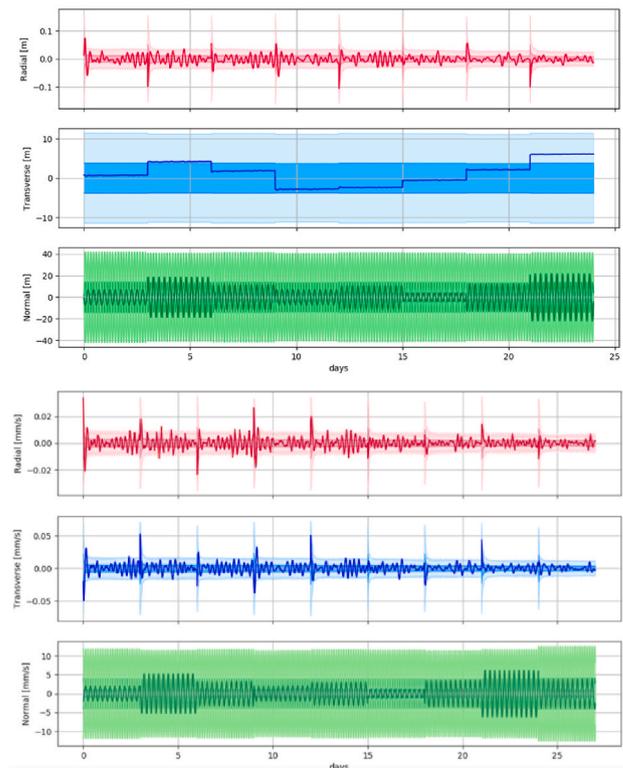

*b) Batch filter with overlaps*

**Fig. 10.** Estimation error (solid lines) with 1-σ and 3-σ uncertainties (shaded area) of RTN components, in case of ISL-only, for constellation B, with a batch sequential filter (a) and with a batch filter with overlap (b).

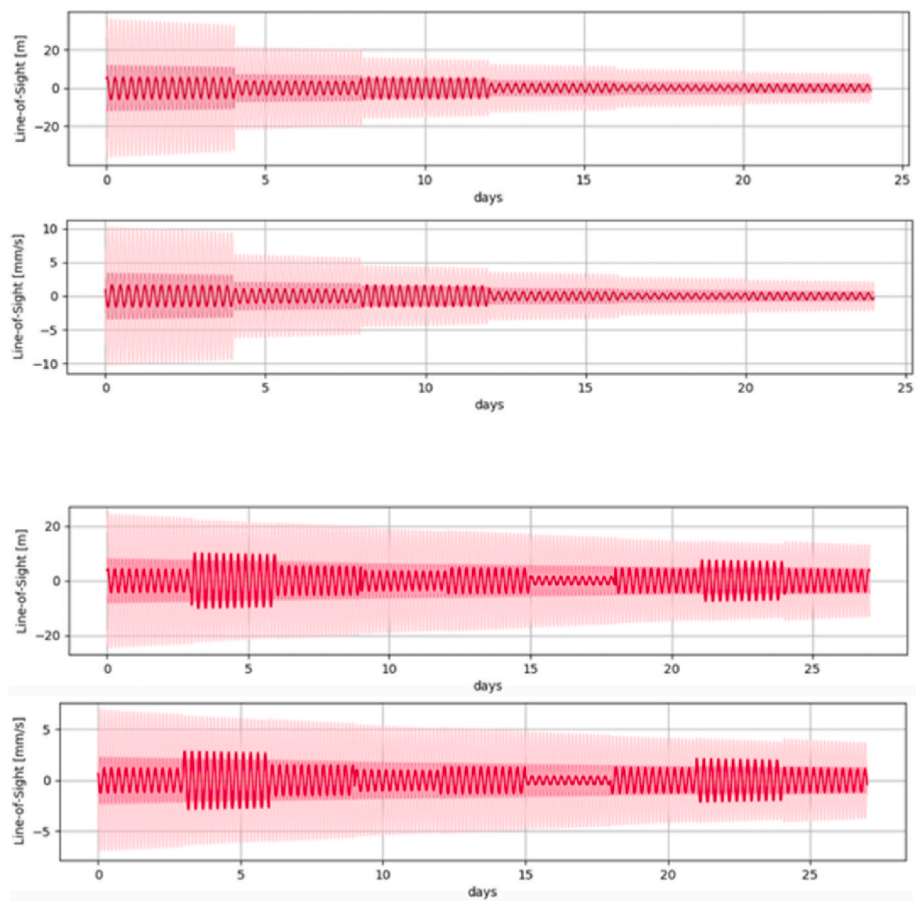

**Fig. 11.** Estimation errors (solid lines) with 1-σ and 3-σ uncertainties (shaded areas) along the line of sight (top: position, bottom: velocity). Results are obtained in the case of ISL-only, with a batch sequential filter (top) and with a batch filter with overlap (bottom).

Fig. 10, with both filters, we obtain a positioning error of ∼500 m on the normal direction at the first arc, and it's not improved by the following additional Doppler data.

The large discrepancy on the normal component can be explained by considering that the ISLs between the nodes of the constellation provide information only on the relative motion between the spacecraft, hence a translation of the entire constellation in a direction orthogonal to the radial does not change remarkably the relative Doppler and range between the nodes.

Furthermore, in case of ISL-only, the orbit determination process relies on gravity gradients, which are well known from the already acquired knowledge of the Mars gravity field and rotational state. Considering that at large distances from Mars the effect of the gravity gradients is negligible, this constellation requires ground support to determine its orbits.

Consequently, in this case, the covariance matrix is underestimated; an effect that is more influent with the use of the batch-sequential filter, where the covariance is mapped to extract the a priori values of the subsequent arcs.

The constellation B, on the contrary, can autonomously reconstruct its position with higher stability, thanks to the dominance of orbit accelerations coming from Mars' gravity field over the solar radiation pressure and other perturbation effects, due to the lower altitude of the orbits.

As shown in Fig. 10, using the batch-sequential filter, after a warm-up period of about 15 days, we obtain a positioning accuracy of ∼10 m, gradually improving as additional Doppler data are analysed with the filter. On the contrary, using the simple batch filter with overlaps, the 3-σ accuracy do not improve over time, since the filter is reinitialized without information from previous data arcs. On the normal direction, the accuracy remains steadily around 40 m.

For both estimation filters, the radial component is the one better determined, with accuracies in the order of 4 cm, stable over time. Also, the estimation error (solid lines) always remains inside the quoted uncertainty envelopes, indicating the goodness of the dynamical model (and thus of the selection of empirical accelerations).

For what concerns the absolute velocity, the Radial and Transverse directions are very well determined (below 0.02 mm/s), while the Normal direction proves again to be the weak direction, with an uncertainty of 5 mm/s and 10 mm/s for the two different estimation filters. If ground support is included (see next paragraph), performances can be increased by more than an order of magnitude. Fig. 11 reports the estimation error and corresponding uncertainty for the relative position and relative velocity between the MSC and one of the DSC projected on the line-of-sight. The relative positioning accuracy is in the order of 5 m and 1 mm/s on average between the nodes. To conclude, by using a batch-sequential filter, constellation B can autonomously reconstruct its trajectory with high accuracy and stability, thanks to the dominance of orbit accelerations coming from Mars' gravity field over the solar radiation pressure, due to the low altitude of the orbits. Mismodelling in Mars gravity and non-gravitational accelerations are correctly accounted for by means of the set of empirical accelerations that have been included in the dynamical model. This can be deduced by the estimation error being always lower than the 3-σ uncertainties.

## 8. Quasi-autonomous system results

In this Section, the orbit determination is performed in a case of a quasi-autonomous system, realized alternating the ISL with ground-tracking or with the use of pseudolites on the Martian surface.

### 8.1. Positioning performance with alternating combination of ISL and ground-tracking

To assess the duration of the Earth-link required by the constellation A to determine the orbits of the nodes, we performed the perturbation analysis simulating an increasing time of ground tracking support, established alternatively with the ISL. We chose the batch-sequential filter for its better performances with respect to the batch filter with overlap, as demonstrated in Section 6.1. In Fig. 12, we illustrate the position error and its relative formal uncertainty obtained after 24 days of simulations in different ground-tracking scenarios. Even 2 h of Earth tracking every 6 days can substantially improve the positioning performance of the constellations, reaching precisions of the order of 10 m. Note that augmenting the ground link duration improves the knowledge on the normal direction and grants an absolute reference point to the constellations, but at the same time decreases the amount of ISL data (the links cannot be simultaneous). This diminution, after a certain threshold, is disadvantageous because the ISL is far more accurate than the Earth-link. So, for this constellation, the best compromise is to alternate.

ISL with ground link in a range between 1 h and 4 h every 3 days.

The same analysis has been carried out for the constellation B, showing performance of the order of ~10 cm with 2 h every 6 days of ground-tracking. The best positioning performances are achieved with 4 h every 3 days of ground-tracking, reaching ~10 cm of accuracy.

### 8.2. Positioning performance with alternating combination of ISL and pseudolites

Implementing the capability of establishing a radio-link from well-known positions of the Mars surface would allow to drastically augment the positioning performances of the constellations. Pseudolites can provide reference points for the orbit determination, serving a similar role as Earth ground stations. The pseudolite is equipped with a transponder that receives and coherently re-transmits back a Doppler tracking signal. This scheme allows establishing two-way coherent links between the MSC and the pseudolites on the Martian surface.

After an initial step in which the positions of the pseudolites have been determined to high accuracy (e.g., with radio links to ground), the location of the pseudolites can be considered as well known and used as fixed points on Mars surface to aid the navigation of the constellation.

The pseudolites shall be deployed on the Mars surface either as independent units requiring a dedicated mission or, leveraging the low mass and power requirements, as a piggyback of other missions (e.g., commercial/scientific landers or rovers, surface hubs).

Having pseudolites on the Mars surface is surely a more expensive solution than establishing an Earth-link, but it's important to assess the beneficial effect on the orbital determination performances.

In these simulations, we considered a link between the MSC and three pseudolites equally spaced on the equator of Mars. For both constellations, the accuracy of the positioning reaches the level of tens of cm, improving of one order of magnitude for constellation A, while remaining approximately at the same level for constellation B, which can be determined, anyhow, more precisely, as shown in Fig. 13.

## 9. Conclusions

We proposed a satellite navigation system able to determine, in almost complete or full autonomy, depending on the altitude of the orbits, the ephemeris of the constellation and we analysed the positioning performance in terms of satellites absolute positioning with respect to Mars. This system exploits a Martian constellation of 5 small satellites on three quasi-circular high-altitude polar orbits, connected via microwave ISL that generate high accuracy Doppler observables.

This navigation system is capable of effectively supporting the new missions planned or proposed in the coming years to explore the Martian polar caps. It could be used in the future for supporting the navigation of other probes, such as rovers or landers during the EDL phase.

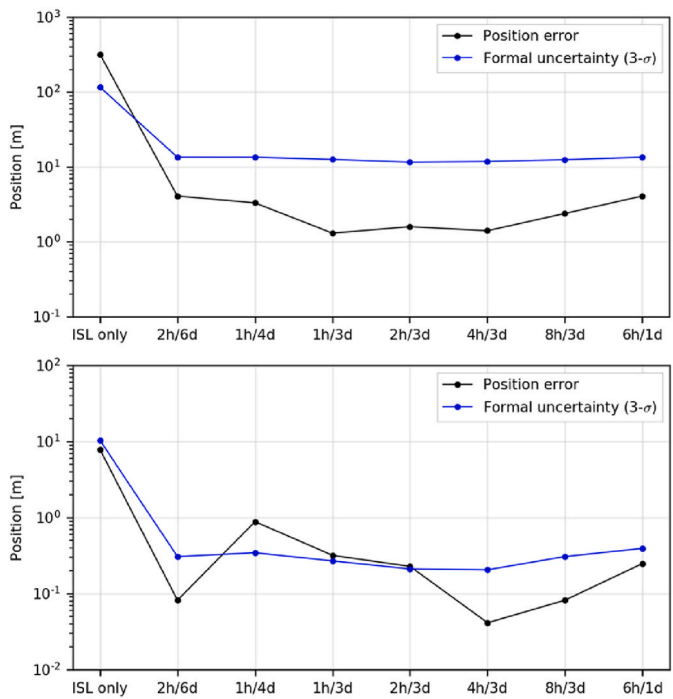

**Fig. 12.** Estimation error (black) and formal uncertainty (blue) after 24 days of simulation with batch-sequential filter for different duration of Earth link, in case of constellation A (top) and constellation B (bottom). (For interpretation of the references to colour in this figure legend, the reader is referred to the Web version of this article.)

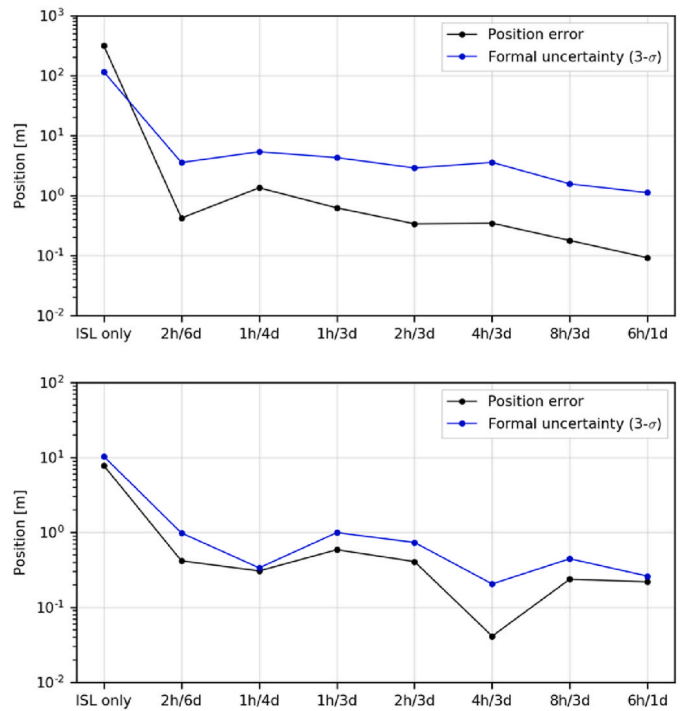

**Fig. 13.** Estimation error (black) and formal uncertainty (blue) after 24 days of simulation with batch-sequential filter for different duration of pseudolites link, in case of constellation A (top) and B (bottom). (For interpretation of the references to colour in this figure legend, the reader is referred to the Web version of this article.)

In fact, although this work does not address yet the positional accuracy of the end user, our analysis shows that the pillar of every satellite navigation system, i.e., precise OD, can be attained by means of a relatively simple radio system configuration, fully compatible with the deployment of a SmallSat constellation. Chronization requirements, techniques and performance with a terrestrial time and within the constellation are presented in [53]. Future works will assess the performance achievable with this concept for spacecraft time synchronization.

The constellations orbits have been chosen to grant coverage of the polar caps and ground-tracks repetitiveness of two days and one day, respectively.

Numerical simulations of the OD process were carried out to investigate the navigation performances of this novel concept. A perturbation analysis, carried out for an autonomous ISL-only case by introducing a 10% mismodeling error on the largest non-gravitational acceleration, shows a positioning accuracy of ~10 m after a warm-up period of about 15 days, if the constellation is sufficiently close to Mars, as our simulations show in the case of constellation B.

With higher semi-major axes, a limited but periodic ground-tracking is required. We show that just 2 h every 6 days can be sufficient to obtain ~10 m positioning accuracy. The best performances are achieved by the lower constellation with 4 h every 3 days of ground-tracking, with accuracies down to ~10 cm. To obtain the same results with the higher constellation, we propose a configuration of 3 pseudolites equally spaced on the Mars equator. This solution, while quite more expensive, would greatly strengthen every orbital solution and provide a strong tie between the constellation and the Mars-centered, Mars-fixed reference frame. The pseudolites would also serve a future, global navigation system. The numerical simulations reported in this study were conducted by analysing deep-space, Mars pseudolites and intersatellite data both with a batch-sequential least-squares filter and a batch filter with overlap, showing that the first one is the preferred method in terms of achievable positioning performances.

The main strength of this novel technology with respect to a GRACE/GRAIL radio link configuration is a significant reduction of the required spacecraft mass and power, enabling the use of small satellites, while still providing highly accurate radio tracking measurements.

## Declaration of competing interest

The authors declare that they have no known competing financial interests or personal relationships that could have appeared to influence the work reported in this paper.

## Acknowledgments

Part of this work was supported by a scientific agreement among University of Rome Sapienza, Italian National Research Council (CNR), and Argotec s. r.l. for a Ph.D. doctorate in aerospace engineering.

## References

[1] B.D. Tapley, et al., The gravity recovery and climate experiment: mission overview and early results, Geophys. Res. Lett. 31 (2004), L09607.

[2] S.W. Asmar, et al., in: GRAIL: Mapping the Moon's Interior, Springer, 2013, pp. 25–55.

[3] M. Di Benedetto, et al., Augmenting NASA europa clipper by a small probe: europa tomography probe (ETP) mission concept, Acta Astronaut. 165 (Dec. 2019) 211–218.

[4] Di Benedetto M. et al ., "Augmenting NASA europa clipper by a small probe: europa tomography probe (ETP) mission concept," in 67th International Astronautical Congress (IAC), Guadalajara, Mexico, 26-30 September 2016.

[5] A.S. Konopliv, An improved JPL Mars gravity field and orientation from Mars orbiter and lander tracking data, Icarus 274 (2016) 253–260.

[6] A. Genova, et al., Seasonal and static gravity field of Mars from MGS, Mars Odyssey and MRO radio science, Icarus 272 (2016) 228–245.

[7] N. Thomas, et al., Mars and the ESA Science Programme-The Case for Mars Polar Science, Experimental astronomy, 2021, pp. 1–17.

[8] R.J. Lillis, et al., Mars Orbiters for Surface-Atmosphere-Ionosphere Connections (MOSAIC), AGU Fall Meeting, 2020.

[9] D.E. Smith, et al., Trilogy, a planetary geodesy mission concept for measuring the expansion of the solar system, Planet. Space Sci. 153 (April 2018) 127–133.

[10] G. Cascioli, et al., The Contribution of a Large Baseline Intersatellite Link to Relativistic Metrology, IEEE 5th International Workshop on Metrology for AeroSpace (MetroAeroSpace), 2019, pp. 579–583, 2019.

[11] S.W. Asmar, et al., Small Spacecraft Constellation Concept for Mars Atmospheric Radio Occultations, American Geophysical Union, Fall Meeting, 2017.

[12] A. Genova, ORACLE: a mission concept to study Mars' climate, surface and interior, Acta Astronaut. 166 (2020) 317–329. January.

[13] S. Goossens, et al., High-resolution gravity field models from GRAIL data and implications for models of the density structure of the Moon's crust, J. Geophys. Res.: Planets 125 (2020).

[14] J. Kim, et al., Simulation of dual one-way ranging measurements, J. Spacecraft Rockets 40 (3) (2003) 419–425.

[15] F.G. Lemoine, et al., High-degree gravity models from GRAIL primary mission data, J. Geophys. Res.: Planets 118 (8) (2013) 1676–1698.

[16] V. di Tana, et al., ArgoMoon: there is a nano-eyewitness on the SLS, IEEE Aero. Electron. Syst. Mag. 34 (4) (2019) 30–36, 1 April.

[17] V. di Tana, et al., LICIACube, the Italian Witness of DART Impact on Didymos, IEEE 5th International Workshop on Metrology for AeroSpace (MetroAeroSpace), 2019, pp. 314–317.

[18] Microsemi, 9700, Ultra-miniature space-qualified OCXO series. https://www.microsemi.com/document-portal/doc_download/133351-9700-datasheet. (Accessed 25 November 2022).

[19] Final Report and Final Presentation of the ESA Study Nr. 4000117076/16/NL/FE on "CDMA Application for TT&C and Precision Navigation"".

[20] Frequency assignment guidelines for communications in the Mars region, Rec. SFCG 22–1R4 (2021). December.

[21] M.M. Kobayashi, Iris deep-space transponder for SLS EM-1 CubeSat missions, in: Small Satellite Conference, 2017.

[23] M.M. Kobayashi, et al., The Iris deep-space transponder for the SLS EM-1 secondary payloads, IEEE Aero. Electron. Syst. Mag. 34 (9) (2019) 34–44. Sept.

[24] MarCO, Mars cube one) [Online]. Available: https://directory.eoportal.org/web/eoportal/satellite-missions/m/marco.

[25] Advanced CubeSat antennas for deep space and Earth science missions: a review [Online]. Available: https://pureadmin.qub.ac.uk/ws/files/174234474/IEEE_Magazine.pdf.

[26] Integrated solar array and reflectarray antenna (ISARA) [Online]. Available: https://www.nasa.gov/directorates/spacetech/small_spacecraft/isara_project.html.

[27] A. Genova, et al., Seasonal and static gravity field of Mars from MGS, Mars Odyssey and MRO radio science, Icarus 272 (2016) 228–245.

[28] D. Durante, et al., Jupiter's gravity field halfway through the Juno mission, Geophys. Res. Lett. 47 (4) (2020).

[29] P. Cappuccio, et al., Report on first inflight data of BepiColombo's mercury orbiter radio science experiment, IEEE Trans. Aero. Electron. Syst. 56 (6) (2020) 4984–4988.

[30] L. Iess, et al., Astra: interdisciplinary study on enhancement of the end- to-end accuracy for spacecraft tracking techniques, Acta Astronaut. 94 (2) (2014) 699–707.

[31] G.M. Resch, et al., Radiometric correction of atmospheric path length fluctuations in interferometric experiments, Radio Sci. 19 (1984) 411–422.

[32] Technical note: allan deviation analysis for ESA DSA2 X/Ka antenna, SED Systems Revision 1 (2003). December.

[33] P.W. Kinman, Doppler tracking of planetary spacecraft, IEEE Trans. Microw. Theor. Tech. 40 (6) (1992) 1199–1204.

[34] B. Tapley, et al., Statistical Orbit Determination, Elsevier, 2004.

[35] E.M. Alessi, et al., Desaturation manoeuvres and precise orbit determination for the BepiColombo mission, Mon. Not. Roy. Astron. Soc. 423 (3) (2012) 2270–2278.

[53] Molli, S., Boscagli, G, Di Benedetto, M., Durante, D., Vigna, L., Iess, L., Time transfer and orbit determination for a Martian navigation system based on smallsats, 9th International Workshop on Tracking, Telemetry and Command Systems for Space Applications (TTC) (2022). In press.